\documentclass[aps,prd,showpacs,twocolumn]{revtex4}

\begin{document}


\begin{flushleft}
ADP-03-119/T557 \\
\end{flushleft}

\title{Testing the Nature of the $D_{sJ}^*(2317)^+$ and 
$D_{sJ}(2463)^+$ States Using Radiative Transitions}
 \author{Stephen Godfrey\footnote{Email address: 
godfrey@physics.carleton.ca}}
\affiliation{
Ottawa-Carleton Institute for Physics \\
Department of Physics, Carleton University, Ottawa,  Canada K1S 5B6 \\
{\it and} \\
Special Research Centre for the Subtatomic Structure of Matter\\
University of Adelaide, Adelaide South Australia 5000, Australia }


\date{\today}

\begin{abstract}
The Babar and CLEO collaborations have recently observed states decaying to 
$D_s^+\pi^0$  and $D_s^{*+}\pi^0$ respectively
and suggest the possible explanation that they are the missing P-wave 
$c\bar{s}$ states with $J^P=0^+$ and $1^+$.
In this note we compare the 
properties of the $D_{sJ}^*(2317)^+$ and $D_{sJ}(2463)^+$ states
to those expected of the $c\bar{s}$ $D_{s0}^*$ and $D_{s1}$ states.  
We expect the $D_{s0}^*$ and $D_{s1}$ with the reported masses to be 
extremely narrow, $\Gamma \sim {\cal O}(10\hbox{ keV})$, with 
large branching ratios to $D^*_s\gamma$ for the $D_{s0}^*$
and to $D^*_s\gamma$ and  $D_s\gamma$ for the $D_{s1}$.  
Crucial to this interpretation of the Babar and CLEO 
observations is the measurement of the radiative transitions.
We note that it may be possible to observe the $D_{s1}(2536)$ in 
radiative transitions to the $D_s^*$.
\end{abstract}

\maketitle

\section{Introduction}

Over the last decade there has been considerable progress in our
understanding of mesons, strongly interacting bound states of quarks 
and antiquarks.  Mesons made of one heavy and one light quark have 
played an important role \cite{reviews}.  
However the theoretical predictions have 
not been sufficiently tested by experimental data to say that we 
truly understand the strong interaction.  This situation has recently 
been highlighted by the discovery of a state,  the 
$D_{sJ}^*(2317)^+$,  with mass $2317$~MeV decaying to $D_s^+\pi^0$
by the Babar Collaboration at the Stanford Linear 
Accelerator Center (SLAC) \cite{babar} and a second state,
the $D_{sJ}(2463)^+$, with mass 
2.463~GeV decaying to $D_s^{*+}\pi^0$ by the CLEO Collaboration at 
the Cornell Electron Storage Ring \cite{cleo}.
These states have also been observed by the Belle Collaboration at KEK 
\cite{belle}.
The $D_{sJ}^*(2317)^+$ was observed in the inclusive $D_s^+ \pi^0$ 
invariant mass distribution \cite{babar}.  
The state has natural spin-parity and 
the Babar collaboration suggest it to be a $J^P=0^+$ state based on 
its low mass.  
The quantum numbers of the final state 
indicate that the decay violates isospin conservation.
Babar found no evidence for the decay $D_{sJ}^*(2317)^+\to 
D_s^{*+}\gamma$ or $ D_s^+\gamma\gamma$ and although
they found no evidence for the decay $D_{sJ}^*(2317)^+\to 
D_s^{+}\gamma$
they see a peak near 2.46~GeV in the $D_s^+\pi^0\gamma$ 
mass distribution but do not claim this as evidence for a new state.

The CLEO collaboration subsequently reported on a signal in the 
$D_s^{*+}\pi^0$ channel at a mass of 2.463~GeV which they refer to as 
the $D_{sJ}(2463)^+$ \cite{cleo}.  
Because the $D_{sJ}(2463)$ lies above the 
kinematic threshold to decay to $DK$ but not $D^*K$ the narrow width 
suggests the decay to $DK$ does not occur.  Since angular momentum and 
parity conservation forbids a $1^+$ state from decaying to two 
pseudoscalars, CLEO suggests the compatability of the $D_{sJ}(2463)$ 
with the $J^P=1^+$ hypothesis.  
CLEO puts limits on the widths of the $D_{sJ}^*(2317)^+$ and
$D_{sJ}(2463)^+$  of $\Gamma < 7$~MeV at 90\% C.L. \cite{cleo}.  More 
importantly for the purpose of this analysis, they give limits on 
radiative transitions of the $D_{sJ}^*(2317)^+$ and
$D_{sJ}(2463)^+$ to $D_s^{*+}\gamma$ and $D_s^{+}\gamma$ final states.

The simplest interpretation is to identify these states as the 
missing $j=1/2$ members of the $c\bar{s}$ $L=1$ multiplet where $j=1/2$ 
is the angular momentum of the $s$-quark.
The observation of these states is surprising because they are
narrower than expected,
are observed in isospin violating $D_s\pi^0$ and 
$D_s^*\pi^0$ final states, and are lower in mass than expected by most
(but not all) calculations
\cite{gi85,gk91,ebert97,de01,lahde00,zeng95,gj95,lewis,bali,bardeen}.  
The Babar and CLEO observations have led to conflicting 
interpretations.  Although the observed mass for the $D_{s0}$
candidate 
(the $J^P=0^+$ member of the ground state $L=1$ $c\bar{s}$ multiplet)
 is consistent with 
some predictions of chiral quark models \cite{bardeen,cqm} in which 
broken chiral symmetry views them as the positive-parity 
partners of the $D_s$ and $D_s^*$ states,
it is considerably lower than expected by most quark models
\cite{gi85,gk91,ebert97,de01,lahde00}
and lattice QCD calculations \cite{lewis,bali}. 
This has led to considerable interest 
\cite{cahn,cheng,adam,colangelo} including the 
proposal that the $D_J(2317)^+$ is a multiquark 
state \cite{barnes03,vanbeveren}, possibly a $DK$ molecule analogous 
to the $K\bar{K}$ interpretation of the $f_0(980)$ and $a_0(980)$.

In this letter we confront the 
$c\bar{s}$ $L=1$ $D_{s0}^*$ and $D_{s1}$ interpretations of these 
states with the theoretical expectations for conventional $c\bar{s}$ states. 

\section{Spectroscopy}

Mass predictions are an important test of QCD motivated potential models 
as well as other calculational approaches for hadron spectroscopy
\cite{gi85,gk91,zeng95,ebert97,de01,gj95,lahde00,lewis,bali,cqm}.  
In QCD-motivated potential models the spin-dependent splittings test the 
Lorentz nature of the confining potential with different combinations 
of Lorentz scalar, vector,\ldots interactions
\cite{gi85,gk91,zeng95,ebert97,de01,gj95,lahde00}.  
Furthermore, the observation of 
heavy-light mesons is an important validation of 
heavy quark effective theory \cite{isgur91,eichten93}
and lattice QCD calculations \cite{lewis,bali}.  
In Table I we summarize predictions for the P-wave $c\bar{s}$ states. 
The two $J=1$ states are linear combinations of  $^3P_1$ 
and $^1P_1$ because for unequal mass quarks, $C$ is no longer a good 
quantum number.  We label these as the $D_1^h$ and $D_1^l$.
Most, but not all, models predict the masses of the 
$D^*_{s0}$ 
and the missing $D_1$ state  to be 
substantially higher than the masses reported by Babar \cite{babar}
and CLEO \cite{cleo}. 
Although it is possible that these models need revision it seems 
unlikely that they would disagree with experiment to such a large degree 
given their general success in describing the meson spectrum.  
A more serious problem is the large discrepancy with 
lattice QCD calculations which give
 $M(D^*_{s0})=2499(13)(5)$~MeV and $M(D_{s1})$=2500(16)(2)~MeV
\cite{lewis}.  If the $D_{sJ}^*(2317)^+$ and
$D_{sJ}(2463)^+$ are identified as the missing 
$^3P_0(c\bar{s})$ and $P_1(c\bar{s})$ states it would pose a serious 
challenge for the lattice calculations.

Quark model calculations \cite{gk91} and heavy quark symmetry 
\cite{isgur91} predict that the 4 
$L=1$ $c\bar{s}$ mesons are grouped into two doublets with properties 
characterized by the angular momentum of the lightest quark, $j=1/2$ 
and $j=3/2$.  The $j=3/2$ states are identified with 
the previously observed 
$D_{s1}(2536)^\pm$ and $D_{sJ}(2573)^\pm$ states \cite{pdg} while 
the $j=1/2$ have not previously been observed.  The $j=3/2$ states are 
predicted to be relatively narrow \cite{gk91}, in agreement with experiment 
\cite{pdg}.  
In contrast, assuming the higher masses predicted by the quark model,
the $j=1/2$ states are 
expected to be rather broad, decaying to $DK$ and $D^*K$ respectively 
with large $S$-wave widths. 
The large width is presumed to explain why they have yet to be observed.  
However, if these states are identified with the recently observed 
$D_{sJ}^*(2317)$ and $D_{sJ}(2463)$ states their masses would be below 
the $DK$ and $D^*K$ thresholds 
so that they would be quite narrow, especially for mesons with such 
high mass.

\begin{table}
\caption{Predictions for the $P$-wave $c\bar{s}$ states. The $J=1$ 
states are linear combinations of the $^3P_1$ and $^1P_1$ states.  In 
column 3 we list $D^{h}_1$ which we take to be the higher mass state 
of the two $J=1$ physical states and $D^{l}_1$ the lower. 
PDG refers to the particle data 
group \cite{pdg} and LGT refers to the lattice gauge theory result. }
\begin{tabular}{l l l l l } \hline
Reference & $^3P_0 $ & $D^{h}_1$  & $D^{l}_1$ & $^3P_2$  \\ \hline
Babar \cite{babar} & 2.32 & 	&	&	 \\
CLEO \cite{cleo} &	&	& 2.463 &	\\
PDG \cite{pdg} 	&  	& 2.535 & 	& 2.574  \\ 
GI \cite{gi85,gk91} & 2.48 & 2.56 & 2.55 & 2.59  \\
ZVR  \cite{zeng95} & 2.38 & 2.52 & 2.51 & 2.58  \\
EGF \cite{ebert97} & 2.508 & 2.569 & 2.515 & 2.560  \\
DE  \cite{de01} & 2.487 & 2.605 & 2.535 & 2.581 \\
GJ  \cite{gj95} & 2.388 & 2.536 & 2.521 & 2.573  \\
LNR  \cite{lahde00} & 2.455 & 2.522 & 2.502 & 2.586  \\
LW [LGT] \cite{lewis} & 2.499 & 2.511 & 2.500 & 2.554 \\
GB [LGT] \cite{bali} &	2.437	&	&	& \\
\hline
\end{tabular}
\end{table}

\section{Radiative Transitions}

While masses are one test of models of hadrons, transitions probe the 
internal structure of the state. Comparison between theory and 
experiment of the branching ratios is an important test of any 
assignment for a state. The Babar collaboration observed the 
$D_{sJ}^*(2317)^+$ in the $D_s \pi^0$ final state and report no 
observation of its decay via radiative transitions \cite{babar}.  
The CLEO collaboration put limits on branching ratios 
of various radiative decays of the $D_{sJ}^*(2317)^+$ and $D_{sJ}(2463)^+$
relative to $\Gamma(D_{sJ}^*(2317)^+ \to D_s^+\pi^0)$
and $\Gamma(D_{sJ}(2463)^+ \to D_s^{*+}\pi^0)$ respectively \cite{cleo}.
Because the $D_{sJ}^*(2317)^+$'s mass is below the 
kinematic threshold for the decay $D_{s0}\to DK$,
the only  kinematically allowed strong decay is  $D_{s0} \to D_s \pi^0$. 
Likewise, the  $D_{sJ}(2463)$ is kinematically forbidden to decay to 
its expected dominant decay mode $D_{s1}\to D^*K$ 
so that the $D_{s1}\to D_s^*\pi^0$ is expected to be dominant .
In both cases the decays $D_{s0}^*\to D_s \pi^0$ and $D_{s1}\to 
D_s^*\pi^0$  violate
isospin and are expected to have quite small partial widths.
Thus,  the 
radiative transitions $D_{s0}^* \to D_s^* \gamma$,
$D_{s1} \to D_s^* \gamma$ and $D_{s1} \to D_s \gamma$ would be expected 
to have prominent branching ratios.  

The $E1$ radiative transitions are given by 
\begin{eqnarray} 
\Gamma(i & \to &  f + \gamma)  \\
& = & \frac{4}{27} \alpha \; 
\langle e_Q \rangle^2 \; 
\omega^3 \;(2J_f +1) \; |\langle ^{2s+1}S_{J'} | r | ^{2s+1}P_J \rangle |^2 
\;{\cal S}_{if} \nonumber
\end{eqnarray}
where ${\cal S}_{if}$ is a statistical factor with ${\cal S}_{if}=1$
for the transitions between spin-triplet states,
$D_{sJ}^{(*)}(1P) \to D_s^*\gamma$ and $D_s^*(2S) \to D_{sJ}(1P) \gamma$,
and ${\cal S}_{if}=3$  for the transition between spin-singlet states,
$D_{s1}\to D_s\gamma$,
$\langle e_Q \rangle$ is an effective quark charge given 
by 
\begin{equation}
\langle e_Q \rangle = {{m_s e_c - m_c e_{\bar{s}}}\over{m_c+m_s}}
\end{equation}
where $e_c= 2/3$ and $e_{\bar{s}}=1/3$ are the charges of the $c$-quark and 
$s$-antiquark given in units of $|e|$, $m_c=1.628$~GeV, 
$m_s=0.419$~GeV are the mass of the $c$ and $s$ quarks taken from Ref. 
\cite{gi85},  $\alpha = 1/137.036$ is the fine-structure constant,
and $\omega$ is the 
photon's energy.  The matrix elements $\langle S | r | P \rangle$, 
given in Table II, were evaluated using the wavefunctions of 
Ref. \cite{gi85}.
Relativistic corrections are included in the E1 transition 
via Siegert's theorem \cite{siegert,mcclary,moxhay} 
by including spin dependent interactions in the Hamiltonian used to 
calculate the meson masses and wavefunctions.   To calculate the 
appropriate photon energies the PDG \cite{pdg} values were used for 
observed mesons while the predictions from Ref. \cite{gi85} were used 
for unobserved states with the following modification.  While
splittings between $c\bar{s}$ states predicted by Ref. \cite{gi85}
are in good agreement with experiment the masses are 
slightly higher than observed so to give a more reliable 
estimate of phase space, the masses used in Table II have been adjusted 
lower by 
18~MeV from the predictions of Ref. \cite{gi85}.  For the $D_{s0}^*$ and 
$D_{s1}$ states we give one set of predictions using the Babar and CLEO 
masses and a second set of predictions using the quark model mass
predictions of Ref. \cite{gi85}.

A final subtlety is that 
the $J=1$ states are linear combinations of $^3P_1$ and $^1P_1$ 
because for unequal mass quarks, $C$ is no longer a good quantum 
number.  Thus, 
\begin{eqnarray}
D_{s1}^{3/2} & = & ^1P_1 \cos\theta + ^3P_1 \sin\theta \nonumber \\
D_{s1}^{1/2} & = & - ^1P_1 \sin\theta + ^3P_1 \cos\theta 
\end{eqnarray}
we use $\theta=-38^o$ and the conventions of Ref. \cite{gk91}
in calculating the widths in Table II which include the appropriate 
factors of $\cos^2\theta$ and $\sin^2\theta$ as appropriate.  
The resulting widths are given in Table II.

In addition to the $E1$ transitions the $M1$ transitions $D_{s1}\to 
D^*_{s0} \gamma$ can also take place. However, we found these partial 
widths to be quite small and unlikely to be observable.

\section{Strong Transitions}

The transition $D_{s0}^* \to D_s \pi^0$ is expected 
to be quite small as it violates isospin.
Athough there are a number of theoretical predictions for 
hadronic transitions between quarkonium levels 
\cite{cho94,voloshin,shifman,ko,ky} we know of none for the 
transition $D_{s0}^* \to D_s \pi^0$.  To estimate this partial width
we turn to known transitions and use existing 
theoretical calculations for guidance. 
This approach should at least help 
us gauge the relative importance of this partial width.  The only 
measured transition is $\psi(2S)\to J/\psi(1S) + \pi^0$ with ${\cal 
B}= 9.7\times 10^{-4}$ \cite{pdg} implying $\Gamma (\psi^\prime\to J/\psi 
\pi^0)=0.27$~keV.  A limit exists on the transition $\Upsilon(2S)\to 
\Upsilon(1S) + \pi^0$ of ${\cal B} (\Upsilon(2S)\to \Upsilon(1S)  
\pi^0)<1.1\times 10^{-3}$ 90\% C.L.  implying 
$\Gamma (\Upsilon(2S)\to \Upsilon(1S)  \pi^0) < 0.05$~keV \cite{pdg}.  
The BR for the transition $D_s^*\to D_s +\pi^0$ is $5.8\pm 2.5\%$  
but the total width is not known.  We can estimate the width by using 
the measured branching ratio
${\cal B}(D_s^*\to D_s \gamma)=(94.2\pm 2.5)\%$ with 
a quark model calculation of the radiative transition $D_s^*\to D_s 
\gamma$.  Combining the partial width given by Ref. \cite{gi85} 
of $\Gamma(D_s^*\to D_s \gamma)=0.125$~keV with the measured branching 
ratio
\cite{pdg} gives $\Gamma(D_s^*\to D_s \pi^0) \simeq 7.7$~eV.  For comparison 
Goity and Roberts \cite{goity01} 
obtain $\Gamma(D_s^*\to D_s \gamma)=0.165$~keV giving 
$\Gamma(D_s^*\to D_s \pi^0)=10$~eV (for the $\kappa=0.45$ solution)
and Ebert {\it et al.} \cite{ebert97}
find $\Gamma(D_s^*\to D_s \gamma)=0.19$~keV giving 
$\Gamma(D_s^*\to D_s \pi^0)=12$~eV.  

For our first attempt to estimate $\Gamma(D_{s0}^* \to D_s \pi^0)$
we rescale $\Gamma(D_s^*\to D_s \pi^0)$ assuming 
a $k_\pi^3$ dependence for the partial widths and find 
$\Gamma(D_{s0}^* \to D_s \pi^0)\simeq 2$~keV. One should take this 
estimate with a grain of salt 
as the $D_{s}^* \to D_s \pi^0$ is an $S \to S$
transition with the final states in 
a relative $P$-wave while the $D_{s0}^* \to D_s \pi^0$ transition is 
a $P\to S$ transition with the final states in a relative $S$-wave so 
there are wavefunction effects we have totally ignored in addition to 
a generally cavalier attitude to kinematic factors.  All we have 
attempted to do is establish the order of magnitude.

A more relevant starting point is the transition $h_c(^1P_1) \to 
J/\psi \pi^0$ which is a $P\to S$ spin-flip transition 
which proceeds via the $E1-M1$ interference term in a multipole 
expansion of the gluonic fields, similar to the $^3P_0 \to ^1S_0$ 
transition we are attempting to estimate.  Ko estimates 
$\Gamma(h_c \to J/\psi \pi^0)\simeq 2.5$~keV \cite{ko}.  
This transition is related to the transition 
$\psi' \to h_c \pi^0$ \cite{ko,voloshin} 
for which Ko \cite{ko} obtains 
${\cal B}(\psi' \to h_c \pi^0)= 3 \times 10^{-3}$.  For comparison
Voloshin \cite{voloshin} finds ${\cal B}(\psi' \to h_c \pi^0)= 
10^{-3}$ so that we should assume a factor of 3 in uncertainty.  These 
transitions are proportional to the pion momentum so that by rescaling the 
estimate of $\Gamma(h_c \to J/\psi \pi^0)$ we find
$\Gamma(D_{s0}^* \to D_s \pi^0)\simeq 2$~keV.  There are two important 
uncertainties in this estimate.  The first is that the matrix elements 
are proportional to $\langle S | r | P \rangle$.  Using the 
wavefunctions of Ref. \cite{gi85} we find  
$\langle 1^3P_0 | r | 1^1S_0 \rangle_{cs}/
\langle 1^3S_1 | r | 1^1P_1 \rangle_{cc} = 1.1$.  The second 
uncertainty is that the matrix elements are ${\cal O}(\alpha_s)$ so 
that the ratio of the widths go like 
$(\alpha_s(c\bar{s}))/\alpha_s(c\bar{c}))^2$ which, given that the 
relevant energy scale is the light quark mass, could contribute an
additional factor of 4 in the width.  Given these uncertainties 
we estimate that $\Gamma(D_{s0}^*(2.32) \to D_s \pi^0) \sim 10$~keV.
We expect similar rates for the decays $D_{s1}\to D_s^*\pi^0$ 
\cite{ds1}.
In addition to the one-pion decay modes,
the $D_{s1}$ state can decay via two-pion 
transitions to the $D_s$ state.  (The decay $D^*_{s0}\to D_s \pi\pi$ 
is forbidden by parity conservation.)  Using Ko's estimate of the 
ratio $\Gamma(h_c \to J/\psi +\pi\pi)/\Gamma(h_c \to J/\psi + 
\pi^0)\simeq 0.16$ \cite{ko} we estimate 
$\Gamma(D_{s1}\to D_s \pi\pi)\simeq 1.6$~keV.
The resulting partial widths and branching ratios are summarized in 
Table II.  

For comparison we also include in Table II the partial widths and 
branching ratios expected for the $1^3P_0(c\bar{s})$ state with mass 
2.466~MeV and the $D_{s1}^{1/2}$ state with mass 2.536~MeV.
The dominant decays for these masses are $D_{s0}^*\to DK$
and $D_{s1}^{1/2}\to D^*K$ with large partial widths.  
Although there is considerable uncertainty in the 
estimate of these widths \cite{gk91,bg} we do expect the $D_{s0}^*$ and 
$D_{s1}^{1/2}$ states with these masses 
to be rather broad with  small branching ratios 
for the radiative transition.  These decays are S-wave so the widths scales 
linearly with the decay products momentum.

For completeness we also include in Table II
other E1 transitions involving the $c\bar{s}$ P-wave states.  
We note that the  $D_{s1}(2536)^\pm$ should 
have a relatively large branching ratio for  its radiative transition 
to $D_s^{*\pm}\gamma$ so that it may be possible to observe 
the $D_{s1}(2536)^\pm$ in this mode.

\begin{table*}
\caption{Predictions for partial widths and branching ratios for 
E1 transitions  $2S\to 1P$ and $1P\to 2S$  and strong decays
in the $D_s$ meson sector. For the $D_{s0}^*$ and $D_{s1}^{1/2}$ 
states we show results for two sets of assumptions.  In the first we 
associate the newly observed $D_{sJ}^*(2.317)$ and $D_{sJ}(2.463)$ 
with the $D_{s0}^*$ and $D_{s1}^{1/2}$ while in the second we 
show partial widths using the quark model predictions for these 
states's masses.  For decays involving the $D_{s1}$ states we include the 
appropriate $\cos^2\theta$ and $\sin^2\theta$ factors corresponding to 
eqn. 3 in the partial widths.
The widths are given in keV unless otherwise noted.  
The masses come from the PDG \cite{pdg} unless otherwise noted.}
\begin{ruledtabular}
\begin{tabular}{l l c c c c c c } 
Initial & Final & $M_i$ & $M_f$ &  $k$ & 
	$\langle 1P | r | nS \rangle $ &  Width  & BR  \\
state  & state & (GeV) & (GeV) & (MeV) & (GeV$^{-1}$) & (keV) & \\
\hline 
$D_{s0}^*(2317)^+$ & $ D_s^* \gamma$ & 
	2.317\footnotemark[1] & 2.112 & 196& 2.17\footnotemark[2]
			 & 1.9 & $\sim 16$ \%  \\
	& $ D_s \pi^0$ & 2.317\footnotemark[1] & 1.968 & 297 &  
	& $\sim$ 10 & $\sim 84\%$  \\
$D_{s0}^*(2466)^+$ & $ D_s^* \gamma$ & 
	2.466\footnotemark[3] & 2.112  & 329 & 2.17\footnotemark[2]
			 & 9.0  & $3\times 10^{-5}$ \\
	& $ DK$ & 2.466\footnotemark[3] &  & 289 & & 
	280 MeV\footnotemark[4] & $\sim 100 \%$  \\
\hline 
$D_{s1}^{1/2}(2.463)$ & $ D_s^* \gamma$ & 
	2.463\footnotemark[5] & 2.112 &  326 & 2.18\footnotemark[2] & 
		5.5 &  24\% \\
	& $ D_s \pi^0$ & 2.463\footnotemark[5] & 1.968 & 297 &  
		& $\sim 10 $		& 43\% \\
	& $ D_s \pi\pi$ & 2.463\footnotemark[5] & 1.968 & 297 & 
		 & $\sim 1.6$ & 7\% \\
	& $D_s \gamma$ & 2.463\footnotemark[5]	
		& 1.968 & 445  & 1.86\footnotemark[2] & 6.2 &	27\% \\
$D_{s1}^{1/2}(2.536)$ & $ D_s^* \gamma$ & 
	2.536\footnotemark[3] & 2.112 &  388 & 2.18\footnotemark[2] & 
	9.2 &  $7\times 10^{-5}$ \\
	   & $ D^*K$ & 2.536\footnotemark[3] & & 384 &   & 130 MeV\footnotemark[4] 
							& $\sim 100\%$  \\
	& $D_s \gamma$ & 2.536\footnotemark[3]	& 1.968 & 504  & 
		1.86\footnotemark[2] & 9.0 &
		$7\times 10^{-5}$ \\
\hline 
$D_{s2}^*$ & $ D_s^* \gamma$ & 
	2.574 & 2.112 & 420 & 2.17\footnotemark[2] & 19 
	& $\sim 1.3 \times 10^{-3}$ \footnotemark[6]  \\
\hline
$D_{s1}^{3/2}$ & $ D_s^* \gamma$ & 
	2.535 & 2.112 &  388 & 2.18\footnotemark[2] & 5.6 & 1.6\%   \\
	& $ D^*K$ & 2.535 & & 382 &   & 340\footnotemark[7] 
							& 97 \%  \\
	& $D_s \gamma$ & 2.535 & 1.968 & 503  & 1.86\footnotemark[2] 
		& 15  &	4.2\%\\
\hline 
$D^*(2S)$ & $ D_{s2}^* \gamma$ & 
	2.714\footnotemark[3] & 2.574 & 136 & 2.60\footnotemark[2] & 1.5 &   \\
	& $ D_{s1}^{3/2} \gamma$ & 
	2.714\footnotemark[3] & 2.535 & 173 & 2.25\footnotemark[2] & 0.5 &    \\
	& $ D_{s1}^{1/2}(2.536) \gamma$ & 
	2.714\footnotemark[3] & 2.536\footnotemark[3]
	 & 172 &2.25\footnotemark[2] & 0.9 &    \\
	& $ D_{s1}^{1/2}(2.463) \gamma$ & 
	2.714\footnotemark[3] & 2.463\footnotemark[5]
	 & 239 &2.25\footnotemark[2] & 2.3 &    \\
	& $ D_{s0}^* \gamma$ & 
	2.714\footnotemark[3] & 2.466\footnotemark[3] & 237 & 1.95\footnotemark[2] & 0.9 &  \\
	 & $ D_{s0}^* \gamma$ & 
	2.714\footnotemark[3] & 2.317\footnotemark[1] & 368 
		& 1.95\footnotemark[2] & 3.4 &  \\
\end{tabular}
\end{ruledtabular}
\footnotetext[1]{From Babar Ref.\cite{babar}.}
\footnotetext[2]{Obtained using the wavefunctions generated from 
	Ref. \cite{gi85}}
\footnotetext[3]{Masses taken from Ref.\cite{gi85} with the 
modification that the predictions have been adjusted 
downward by 18~MeV to give better agreement with the 
measured masses.  The masses in Ref. \cite{gi85} were rounded to 
10~MeV.  Here we round to 1~MeV.}
\footnotetext[4]{Obtained by rescaling the result of Ref. \cite{gk91} 
by phase space.}
\footnotetext[5]{From CLEO Ref.\cite{cleo}.}
\footnotetext[6]{Based on the PDG total width for the $D_{sJ}(2573)^\pm$ \cite{pdg}}
\footnotetext[7]{The PDG gives $\Gamma < 2.3$~MeV 90\% C.L.. 
We used the width given by Ref. \cite{gk91} rescaled for phase space.}
\end{table*}

CLEO \cite{cleo} has obtained 90\% C.L. 
limits on radiative transitions of the 
$D_{sJ}^*(2317)$ and $D_{sJ}(2463)$ which we summarize along with our 
predictions in Table III.

\begin{table}
\caption{Comparison of 90\% C.L. limits on radiative transitions 
obtained by CLEO \cite{cleo} with the predictions given in Table 
II.  The BR's are with respect to the decay $D^*_{s0}(2317)\to D_s 
\pi^0$ for the $D^*_{sJ}(2317)$ and with respect to the decay 
$D_{s1}^{1/2}(2463)\to D_s^* \pi^0$  for the $D_{sJ}(2463)$.}
\begin{tabular}{l l l  } \hline
Transition  & Predicted   & CLEO \cite{cleo} \\ \hline
$D^*_{sJ}(2317)\to D_s^{*+} \gamma$ & 0.19 & $< 0.059$ \\
$D^*_{sJ}(2317)\to D_s^+ \gamma$ & 0 & $< 0.052$ \\
$D_{sJ}(2463)\to D_s^{*+} \gamma$ & 0.55 & $< 0.16$ \\
$D_{sJ}(2463)\to D_s^+ \gamma$ & 0.62 & $< 0.49$ \\
$D_{sJ}(2463)\to D^*_{sJ}(2317) \gamma$ & $1.2\times 10^{-3}$ & $<0.58$ \\
\hline
\end{tabular}
\end{table}

\section{Other Possibilities}

If the radiative transitions are not observed with BR's consistent with 
those of the conventional $D_{s0}^*$ and $D_{s1}$ states what are the alternatives? 
One possibility suggested by the Babar collaboration 
is that the  $D_{sJ}^*(2317)^+$ is some sort of 
multiquark state, either a $DK$ molecule or a $c\bar{q}q\bar{s}$ 
multiquark object.  This seems to be a likely possibility which has 
much in common with the description of the $f_0(980)$ and $a_0(980)$ 
as multiquark states: The $D_{sJ}^*(2317)^+$ lies just below the 
$DK$ threshold while the $f_0(980)/a_0(980)$ lie just below the 
$K\bar{K}$ threshold and both couple strongly to these nearby channels.
This explanation has been promoted by Barnes, Close and 
Lipkin \cite{barnes03} and is supported by a recent dynamical 
calculation by van Beveren and Rupp \cite{vanbeveren}. Likewise, the 
$D_{sJ}(2463)$ could be a $D^*K$ bound state similar to the 
$K^*\bar{K}$ molecule interpretation advocated as the solution to the 
longstanding $E/\iota$ puzzle \cite{gn}.

\section{Conclusions}

The discovery of the $D_{sJ}^*(2317)^+$ and $D_{sJ}(2463)$ 
has presented an interesting 
puzzle to meson spectroscopists.  The Babar and CLEO collaborations believe 
that they may be the missing $J^P=0^+$ and $1^+$ members 
of the $L=1 (c\bar{s})$ multiplet.
However their masses are significantly lower than 
expected by most models and also lattice QCD calculations
and would pose a serious 
challenge to these calculations.  It is therefore important to test 
these assignments.   
If the $D_{sJ}^*(2317)^+$ and $D_{sJ}(2463)$ 
are conventional $D_{s0}^*$ and $D_{s1}^{1/2}$ $(c\bar{s})$ states
we have argued that they 
should have very small total widths, ${\cal O}(10)$~keV, with  large 
branching ratios to $D^*_s \gamma$ (and $D_s\gamma$ for the $D_{s1}^{1/2}$).  
It is therefore important to make 
a better determination of the total width of these states and to search 
for the radiative transitions.  In contrast, the absence of the 
radiative transitions and a relatively large total
width  of ${\cal O}(\hbox{MeV})$ would support the 
$D^{(*)}K$ molecule designations.   In this case the conventional 
$D_{s0}^*$ and $D_1^{1/2}$ states have yet to be discovered, 
presumably due to their large width.  However, observation of their 
non-strange partners by the Belle collaboration \cite{belle02} with 
their expected properties 
leads us to be hopeful that they can be found.

\acknowledgments

The author thanks Ted Barnes and Sheldon Stone for useful comments and 
suggestions and Frank Close, Randy Lewis, and Jon Rosner
for helpful communications.
This research was supported in part by the Natural Sciences and Engineering 
Research Council of Canada.




\begin{thebibliography}{99}


\bibitem{reviews}
J.~Bartelt and S.~Shukla,
Ann.\ Rev.\ Nucl.\ Part.\ Sci.\  {\bf 45}, 133 (1995).


\bibitem{babar}
B.~Aubert {\it et al.}  [BABAR Collaboration],
arXiv:hep-ex/0304021.

\bibitem{cleo}
D.~Besson {\it et al.} [CLEO Collaboration],
arXiv:hep-ex/0305100.

\bibitem{belle}
T.Browder [Belle Collaboration], talk given at the 
8th Conference on the Intersections of Particle and Nuclear Physics 
     (19 - 24, May 2003, New York, USA).

\bibitem{gi85}
S.~Godfrey and N.~Isgur,
Phys.\ Rev.\ D {\bf 32} (1985) 189.

\bibitem{gk91}
S.~Godfrey and R.~Kokoski,
Phys.\ Rev.\ D {\bf 43}, 1679 (1991).

\bibitem{zeng95}
J.~Zeng, J.~W.~Van Orden and W.~Roberts,
Phys.\ Rev.\ D {\bf 52}, 5229 (1995)
[arXiv:hep-ph/9412269].

\bibitem{ebert97}
D.~Ebert, V.~O.~Galkin and R.~N.~Faustov,
Phys.\ Rev.\ D {\bf 57}, 5663 (1998)
[Erratum-ibid.\ D {\bf 59}, 019902 (1999)]
[arXiv:hep-ph/9712318].

\bibitem{de01}
M.~Di Pierro and E.~Eichten,
Phys.\ Rev.\ D {\bf 64}, 114004 (2001)
[arXiv:hep-ph/0104208].

\bibitem{gj95}
S.~N.~Gupta and J.~M.~Johnson,
Phys.\ Rev.\ D {\bf 51}, 168 (1995)
[arXiv:hep-ph/9409432].

\bibitem{lahde00}
T.~A.~Lahde, C.~J.~Nyfalt and D.~O.~Riska,
Nucl.\ Phys.\ A {\bf 674}, 141 (2000)
[arXiv:hep-ph/9908485].


\bibitem{lewis}
R.~Lewis and R.~M.~Woloshyn, Phys.\ Rev.\ D {\bf 62}, 114507 (2000);
Nucl.\ Phys.\ Proc.\ Suppl.\ {\bf 94}, 359 (2001).

\bibitem{bali}
G.~S.~Bali,
arXiv:hep-ph/0305209.

\bibitem{bardeen}
W.~A.~Bardeen, E.~J.~Eichten and C.~T.~Hill,
arXiv:hep-ph/0305049.

\bibitem{cqm}
M.~A.~Nowak, M.~Rho and I.~Zahed,
Phys.\ Rev.\ D {\bf 48}, 4370 (1993)
[arXiv:hep-ph/9209272];
M.~A.~Nowak and I.~Zahed,
Phys.\ Rev.\ D {\bf 48} (1993) 356;
W.~A.~Bardeen and C.~T.~Hill,
Phys.\ Rev.\ D {\bf 49}, 409 (1994)
[arXiv:hep-ph/9304265].

\bibitem{cahn}
R.N. Cahn and J.D. Jackson, arXiv:hep-ph/0305012.

\bibitem{cheng}
H.~Y.~Cheng and W.~S.~Hou,
arXiv:hep-ph/0305038.

\bibitem{adam}
A.P. Szczepaniak, arXiv:hep-ph/0305060.

\bibitem{colangelo}
P.~Colangelo and F.~De Fazio,
arXiv:hep-ph/0305140.

\bibitem{barnes03}
T.~Barnes, F.~E.~Close and H.~J.~Lipkin,
arXiv:hep-ph/0305025.

\bibitem{vanbeveren}
E.~van Beveren and G.~Rupp,
arXiv:hep-ph/0305035.

\bibitem{isgur91}
N.~Isgur and M.~B.~Wise,
Phys.\ Rev.\ Lett.\  {\bf 66}, 1130 (1991).

\bibitem{eichten93}
E.~J.~Eichten, C.~T.~Hill and C.~Quigg,
Phys.\ Rev.\ Lett.\  {\bf 71}, 4116 (1993)
[arXiv:hep-ph/9308337].

\bibitem{pdg}
Particle Data Group, 
K. Hagiwara et al, Phys. Rev. {\bf D66}, 010001 (2002).

\bibitem{siegert}
A.~J.~Siegert,
Phys.\ Rev.\  {\bf 52}, 787 (1937).

\bibitem{mcclary}
R.~McClary and N.~Byers,
Phys.\ Rev.\ D {\bf 28}, 1692 (1983).

\bibitem{moxhay}
P.~Moxhay and J.~L.~Rosner,
Phys.\ Rev.\ D {\bf 28}, 1132 (1983).



\bibitem{goity01}
J.L. Goity and W. Roberts, Phys.\ Rev.\ D {\bf 64}, 094007 (2001).

\bibitem{cho94}
P.~L.~Cho and M.~B.~Wise,
Phys.\ Rev.\ D {\bf 49}, 6228 (1994)
[arXiv:hep-ph/9401301].

\bibitem{voloshin} 
M. B. Voloshin, Sov. J. Nucl. Phys. {\bf 43}, 1011 (1986);
M. B. Voloshin and V. I. Zakharov, 
Phys. Rev. Lett. {\bf 45}, 688 (1980).

\bibitem{shifman}
B.L. Ioffe and M.A. Shifman, Phys. Lett. {\bf 95B}, 99 (1980);
V.A. Novikov and M.A. Shifman, Z. Phys. {\bf C8}, 43 (1981);
M.B. Voloshin, hep-ph/0302261.

\bibitem{ky}
Y. P. Kuang and T. M. Yan, Phys. Rev. {\bf D24}, 2874 (1981);
T. M. Yan, Phys. Rev. {\bf D22}, 1652 (1980);
Y. P. Kuang, S. F. Tuan and T. M. Yan, Phys. Rev. {\bf D37}, 1210 (1988).

\bibitem{ko}
P. Ko, Phys. Rev. {\bf D52}, 1710 (1995).

\bibitem{ds1}
See also Ref. \cite{colangelo} and Ref. \cite{bardeen}.

\bibitem{bg}
A discussion about uncertainties in these decay models appears in
H.G. Blundell and S. Godfrey Phys. Rev. {\bf D53}, 3700 (1996).


\bibitem{gn}
For a discussion of this puzzle see S.~Godfrey and J.~Napolitano,
Rev.\ Mod.\ Phys.\  {\bf 71}, 1411 (1999)
[arXiv:hep-ph/9811410].

\bibitem{belle02}
K. Abe {\it et al.}, [Belle Collaboration], contributed paper to the 
XXXI International Conference on High Energy Physics
     (24 - 31 July 2002, Amsterdam, The Netherlands)  BELLE-CONF-0235.


\end{thebibliography}
\end{document}